\documentstyle[12pt,twoside]{article}
\def\greaterthansquiggle{\raise.3ex\hbox{$>$\kern-.75em\lower1ex\hbox{$\sim$}}}
\def\lessthansquiggle{\raise.3ex\hbox{$<$\kern-.75em\lower1ex\hbox{$\sim$}}}
\newcommand{\beq}{\begin{equation}}
\newcommand{\eeq}{\end{equation}}
\newcommand{\beqa}{\begin{eqnarray}}
\newcommand{\eeqa}{\end{eqnarray}}
\newcommand{\beqan}{\begin{eqnarray*}}
\newcommand{\eeqan}{\end{eqnarray*}}
\newcommand{\ba}{\begin{array}}
\newcommand{\ea}{\end{array}}

\newcommand{\A}{{\cal A}}

\newcommand{\Ha}{{\cal H}}

\def\nz{\ifmmode {I\hskip -3pt N} \else {\hbox {$I\hskip -3pt N$}}\fi}
\def\zz{\ifmmode {Z\hskip -4.8pt Z} \else
       {\hbox {$Z\hskip -4.8pt Z$}}\fi}
\def\qz{\ifmmode {Q\hskip -5.0pt\vrule height6.0pt depth 0pt
       \hskip 6pt} \else {\hbox
       {$Q\hskip -5.0pt\vrule height6.0pt depth 0pt\hskip 6pt$}}\fi}
\def\rz{\ifmmode {I\hskip -3pt R} \else {\hbox {$I\hskip -3pt R$}}\fi}
\def\cz{\ifmmode {C\hskip -4.8pt\vrule height5.8pt\hskip 6.3pt} \else
       {\hbox {$C\hskip -4.8pt\vrule height5.8pt\hskip 6.3pt$}}\fi}
\newtheorem{theorem}{Theorem}
\newtheorem{definition}{Definition}
\newtheorem{lemma}{Lemma}

\normalsize
\begin{document}
\bibliographystyle{plain}
\begin{titlepage}
\begin{center}
{\Large \bf Nonequilibrium steady states 
on 1-d lattice systems 
\\ 
and 
Goldstone theorem}\\[24pt]
Takayuki Miyadera \\
Department of Information Sciences \\
Tokyo University of Science \\
Noda City, Chiba 278-8510,
Japan
\vfill
{\bf Abstract} \\
\end{center}
On one-dimensional two-way infinite lattice system,
a property of
stationary (space-) translationally invariant states with nonvanishing 
current expectations are investigated.
We consider GNS representation with respect to such a state,
on which we have a group of space-time translation unitary operators.
We show, by applying Goldstone-theorem-like argument,
that spectrum of the unitary operators, energy-momentum
spectrum with respect to the state, has a singularity at 
the origin.
\vfill
\small
e-mail: miyadera@is.noda.tus.ac.jp
\end{titlepage}
\normalsize
\section{Introduction} 
Recently a lot of researchers get interested in nonequilibrium 
states. Despite their efforts, in contrast to equilibrium state, 
it still has no rigid universal standing point. In equlibrium state 
business, we have KMS (Kubo-Martin-Schwinger) condition to 
be treated, which has been intensively examined to indeed deserve 
the name of equilibrium state \cite{HT,PW,BR}. 
Nonequilibrium state is, by its name, 
state which is not equilibrium state, thus not KMS state. 
It yields too much variety of 
 states to investigate and some restriction should 
be needed to draw any meaningful results. 
The researchers therfore consider
variously restricted situations depending upon their own interests
\cite{HoA,Ta,Shmz,Oji,Fr,AI,SST,JP} and
seek for what physically reasonable non-equilibrium states are.
We, in the present paper, take a slightly different point of view.  
We take a minimal condition to define nonequilibrium 
steady state and discuss its property.
We call the following state a nonequlibrium steady state:
a time-invariant (stationary) and translationally invariant 
state with nonvanishing current
on one-dimensional lattice system.
This restriction indeed is still too weak from the physical point of view,
since it yields physically reasonable staes but also 
yields a lot of unphysical states (states which are hard to realize). 
We apply a Goldstone-theorem-like argument to 
such a nonequilibrium steady
state to show a characteristic behavior of energy momentum 
spectum in its origin. 
We consider GNS representation with respect to 
a nonequilibrium steady state $\omega$, in which
we have a space-time translation unitary operator thanks to 
time and translational invariance.
We show that the spectrum of the unitary operator has 
a singularity at the origin. 
\par
The paper is organized as follows. In next section we 
briefly introduce one-dimensional lattice system and 
define nonequilibrium steady state on it. And in section \ref{sec:gold},
we briefly review Goldstone theorem in nonrelativistic 
setting. Finally we apply the argument to our 
nonequilibrium steady state.
\section{Nonequilibrium Steady State on 1-d lattice system}
 We deal with a quantum one-dimensional two-way 
 infinite lattice system.
To each site $x \in {\bf Z}$ a Hilbert space $\Ha_{x}$ which is 
isomorphic to 
${\bf C}^{N+1}$
is attached  and
the observable algebra at site $x$ is a matrix algebra 
on $\Ha_x$ 
which is denoted by $\A(\{x\})$. 
The observable algebra on 
a finite set 
$\Lambda \subset {\bf Z}$ 
is a matrix algebra 
on $\otimes_{x \in \Lambda} \Ha_x$ and denoted by 
$\A(\Lambda)$. Natural identification can be used to 
derive the inclusion property 
$\A(\Lambda_1) \subset \A(\Lambda_2)$
for $\Lambda_1 \subset \Lambda_2$.
The total observable algebra is a norm completion of 
the sum of the finite rigion observable algebra, 
$\A := \overline{\cup_{\Lambda: \mbox{\small{finite}}} 
\A(\Lambda)}^{\Vert \Vert}$,
which becomes $C^*$ algebra. (For detail, see \cite{BR}.)

To discuss the dynamics, we need a one-parameter 
$*$-automorphism group on $\A$, which we assume 
is induced by a local interaction. That is, there are self-adjoint
operators $\Phi(X) \in \A(X)$ for $d(X):=max\{|x-y|;x,y \in X\} \leq r$ 
(or for $d(X) > r$, $\Phi(X)=0$) and the local Hamiltonian 
of a finite region $\Lambda$ is defined by 
\begin{eqnarray*}
H_{\Lambda} := \sum_{X \subset \Lambda} \Phi(X).
\end{eqnarray*}
The positive integer $r$ is called range of the interaction.
Here we assume translational invariance of 
$\Phi(X)$. That is,
\begin{eqnarray*}
\tau_x(\Phi(X))=\Phi(X+x),
\end{eqnarray*}
holds
for each $x\in {\bf Z}$ 
where $\tau_x$ is a space translation $*$-automorphism.
The Hamiltonian 
defines a one-parameter $*$-automorphism $\alpha_t$ by
\begin{eqnarray*}
\frac{d \alpha_t(A)}{dt}:= 
-i \lim_{\Lambda \to {\bf Z}}[\alpha_t(A), H_{\Lambda}]
\end{eqnarray*} 
for all $A\in \A$. 
\\
To define a current operator we assume the existence of local charge 
operators. 
Namely there exists a self-adjoint operaor $n_x \in \A(\{x\})$
for each $x \in {\bf Z}$ with $\tau_x(n_0)=n_x$ 
and we put $N_{\Lambda}:=\sum_{x \in \Lambda}n_x$.
The local charge defines a one-parameter $*$-automorphism group on the 
observable algebra by
\begin{eqnarray*}
\frac{d \gamma_{\theta}(A)}{d\theta}
=i \lim_{\Lambda \to {\bf Z}}[N_{\Lambda},\gamma_{\theta}(A)].
\end{eqnarray*}
We assume $N_{\Lambda}$ is conserved with respect to 
$H_{\Lambda}$, that is, 
\begin{eqnarray*}
[N_{\Lambda},H_{\Lambda}]=0
\end{eqnarray*}
holds for any finite region $\Lambda$. 
By letting $\Lambda \to {\bf Z}$ this relation derives a purely 
algebraic relation,
\begin{eqnarray*}
\alpha_t\circ \gamma_{\theta}=\gamma_{\theta}\circ \alpha_t.
\end{eqnarray*}
With this relation, $\gamma_{\theta}$ is called a (continuous) 
symmetry transformation.
\\
To define a current operator, we should remind that 
the current is nothing but a charge flow.
If we consider the equation of motion for
the charge contained in a finite region $\Lambda:=[-L,0]$,
 we should obtain, for a sligtly larger region $\Lambda_1 \supset
\Lambda$,
\begin{eqnarray*}
\frac{d \alpha_t(N_{\Lambda})}{dt}|_{t=0}
=-i[N_{\Lambda}, H_{\Lambda_1}]
=j_{-L} -j_0,
\end{eqnarray*}
where $j_{-L}$ represents an in-going charge flow (current) at the
left boundary and $j_0$ represents an out-going 
charge flow (current) at the right boundary.
This equation corresponds to a continuity equation 
in continuum case.
To obtain only the term $j_0$, we deform the 
above equation of motion to pick out the 
right boundary term.    
The current operator at the origin is hence defined by 
\begin{eqnarray*}
j_0:= i [N_{[-L,0]}, H_{[-M,M]}]
\end{eqnarray*}
for sufficiently large $L>0$, $M>0$ and $L-M>0$ 
in comparison to the range of the interaction $r$. 
Note that this current defining
 equation does not depend upon  
the choice of $M$ and $L$ if they satisfy the above conditions.

The above seemingly abstract setting has 
physically interesting examples.
For instance, interacting fermion system can be treated. 
 $\Phi(\{x\})=0$, $\Phi(\{x,x+1\})=
-t (c^*_{x+1}c_{x}+ c^*_x c_{x+1}) +v(1)n_x n_{x+1}$ and 
$\Phi(\{x,x+s\})=v(s)n_x n_{x+s}$ for $2 \leq s \leq r$ gives a 
finite range hamiltonian and 
$n_x:=c^*_x c_x$ is a charge.
The current at the origin is calculated as 
$j_0= it (c^*_1 c_0 - c^*_0 c_1)$.
Heisenberg model can be another example. 
$\Phi(\{x,x+1\}):= S_{x}^{(1)}S_{x+1}^{(1)}
+S_{x}^{(2)}S_{x+1}^{(2)}+\lambda S_{x}^{(3)}S_{x+1}^{(3)}$ 
and $n_x:=S_x^{(3)}$ leads $j_0=-S_0^{(2)}S_1^{(1)}
+S_0^{(1)}S_1^{(2)}$.

Now we introduce the notion of nonequilibrium steady state.
\begin{definition}
A state $\omega$ over two-way 
infinite lattice system $\A$ is called a nonequilibrium steady state 
iff the following conditions are all satisfied:
\begin{item}
\item[(1)] 
$\omega$ is stationary, i.e., $\omega\circ \alpha_t =\omega$ for all $t$. 
\item[(2)]
$\omega$ is translationally invariant. i.e,. $\omega\circ \tau_x =\omega$
for all $x$.
\item[(3)]
$\omega$ gives non-vanishing expectation of the current,
i.e., 
$
\omega(j_0) \neq 0.
$
\end{item}
\end{definition}
Here we do not impose any other condition, stability for instance.
Our definition hence might include rather unphysical states which should be 
hardly realized. 
It, however, contains physically interesting states,
for instance, stationary states obtained by 
inhomogenious initial conditions which was dicussed in \cite{Ta}.
Therfore it is meaningful from the 
physical point of vies to discuss such a state.
\par
We put a GNS represntation with respect to a nonequlibrium state
$\omega$ as $(\Ha, \pi, \Omega)$. Since we fix a state $\omega$, indices
showing the dependence on $\omega$ are omitted.
Moreover we identify $A$ and $\pi(A)$ and omit to write $\pi$.

Since the nonequlibrium steady state $\omega$ is stationary 
and translationally invariant, one can define a 
unitary operator $U(x,t)$ for each $x\in {\bf Z},\ t\in {\bf R}$ 
on ${\cal H}$ by
\begin{eqnarray*}
U(x,t)A\Omega := \alpha_t\circ\tau_x(A)\Omega
\end{eqnarray*}
for each $A \in {\cal A}$. 
Thanks to commutativity of time and space translation,
the unitary operators satisfy
\begin{eqnarray*}
U(x_1,t_1)U(x_2,t_2)=U(x_1+x_2,t_1+t_2)
\end{eqnarray*}
and can be diagonalized into the form:
\begin{eqnarray*}
U(x,t)=\int_{k=-\pi}^{\pi} \int_{\epsilon=-\infty}^{\infty}
e^{i(\epsilon t-kx)}E_{\omega}(dkd\epsilon).
\end{eqnarray*}
The corresponding generator of time translation, $H_\omega$, 
can be written for $A\in \A$ as 
\begin{eqnarray*}
H_\omega A\Omega := \lim_{\Lambda \to {\bf Z}} [H_{\Lambda,}, A]\Omega,
\end{eqnarray*}
whose spectrum decomposition has the form:
\begin{eqnarray*}
H_{\omega}=\int \epsilon E_{\omega}(dk d\epsilon).
\end{eqnarray*}
In the following sections, we investigate the property of 
$E_{\omega}(dkd\epsilon)$.
%
\section{Goldstone theorem}\label{sec:gold}
In this section we give a brief review of
Goldstone theorem \cite{Gold,Ezawa} 
with nonrelativistic setting \cite{LFW}.
The topic was extensively investigated by Requardt \cite{Req}.
In the present paper, we sketch their result without rigor just 
only for giving a hint to our problem. 
\\
Let us consider a $d$-dimensional lattice system 
($d\geq 1$). The dynamics $\alpha_t$ is induced by
a local interaction. Assume there exists a local charge which 
is conserved and induces a continuous 
one parameter automorphism group $\gamma_{\theta}$ on ${\cal A}$. 
A state is said symmetry breaking iff it is not invariant
with respect to $\gamma_\theta$, i.e., $\omega\circ \gamma_{\theta}
\neq \omega$. In the differential form, it is expressed as follows:
there exists a self-adjoint operator $a\in \A$ such that,
\begin{eqnarray*}
i \lim_{\Lambda \to {\bf Z}}(\Omega,[\hat{N}_{\Lambda},\hat{a}]
\Omega)=c \neq 0,
\end{eqnarray*}
where we use the notation 
$\hat{N}_{\Lambda}:=N_{\Lambda}-\omega(N_{\Lambda})$ and
$\hat{a}:=a-\omega(a)$.
When the symmetry breaking state $\omega$ is stationary,
one can easily show the following significant observation;
\begin{eqnarray}
\lim_{\Lambda \to {\bf Z}}(\Omega,i[\hat{N}_{\Lambda},\alpha_t(\hat{a})]
\Omega)=c \mbox{ (time-indep.) }\neq 0.
\label{goldstone}
\end{eqnarray}
Integration of (\ref{goldstone}) with an arbitrary function $f_T$
whose support is included in $[-T,T]$ leads
\begin{eqnarray}
\lim_{\Lambda \to {\bf Z}}
\int dt f_T(t)(\Omega,i[\hat{N}_{\Lambda},\alpha_t(\hat{a})]
\Omega)=\sqrt{2\pi} c\tilde{f}_T(0),
\label{goldstone2}
\end{eqnarray}
where $\tilde{f}_T(\epsilon)=\frac{1}{2\pi}\int dt f(t)e^{i\epsilon t}$
is Fourier transform of $f_T$.
To investigate the property of $E_{\omega}(dkd\epsilon)$, 
we define a {\it function} $\psi(k,\epsilon)$ by
\begin{eqnarray*}
\psi(k,\epsilon)dkd\epsilon
=i(\Omega,\hat{n}E_{\omega}(dkd\epsilon)\hat{a}\Omega),
\end{eqnarray*}
then we obtain
\begin{eqnarray*}
\sqrt{2\pi}2\pi \int
d\epsilon (\psi(0,\epsilon)+\psi^{*}(0,-\epsilon))
\tilde{f}_T(\epsilon)
=\sqrt{2\pi} c\tilde{f}_T(0)
\end{eqnarray*}
Thus we obtain a singularity at the origin:
\begin{eqnarray*}
2\pi(\psi(0,\epsilon)+\psi^*(0,-\epsilon))=c
\delta(\epsilon),
\end{eqnarray*}
where $\delta$ represents Dirac delta function.
This is a nonrigorous sketch of Goldstone theorem.
\par
Note that under rather generic setting one obtains 
${\bf R}$ as specturm of time translation generator, $H_{\omega}$
\cite{HT}.
Therefore the often stated {\it absense of mass gap} gives 
nontrivial information only for the restricted situations like vacuum states.
Manifestation of singularity is the significant result of 
Goldstone theorem.
\par
To show the often claimed {\it poor decay of spatial correlation},
the state in consideration should be a KMS state.
One can utilize Bogoliubov inequality for investigation of spatial 
correlation.
\section{Spectrum in Nonequilibrium Steady State}
Now we go back to 1-d lattice system and a nonequilibrium steady 
state $\omega$ on it.
The following is the main theorem of the present paper.
\begin{theorem}\label{th:th1}
For a nonequilibrium steady state $\omega$, 
$E_{\omega}(dkd\epsilon)$ has a singularity at 
the origin $(k,\epsilon)=(0,0)$.
\end{theorem}
The proof goes similarly to Goldstone theorem and 
the point is to esitimate the following quantity:
\begin{eqnarray*}
\int dt (\Omega, i [N_{[-L,0]}, H_{[-M,M]}(t)] \Omega) f_T(t),
\end{eqnarray*}
where $f_T$ is a real function with 
$\mbox{supp}f_T \subset [-T,T]$. 
Since $(\Omega, i [N_{[-L,0]}, H_{[-M,M]}(t)] \Omega)$
is not time invariant, proof of Goldstone theorem is 
not directly applicable. However, it is almost time invariant 
for sufficiently large $L$ and $M$. 
Wee prove the theorem at the end of this section.
We repeatedly employ the following lemma (see Appendix for proof).
\begin{lemma}\label{th:gp}
Let $V(\Phi)$ be a quantity which is determined by the interaction
$\Phi$ such that
\begin{eqnarray*}
V(\Phi):=sup_{x \in {\bf Z}} \sum_{X \ni x}
|X| (N+1)^{2|X|} e^{r} \Vert\Phi(X)\Vert,
\end{eqnarray*}
where $|X|$ denotes a number of sites included in $X$.
For all $A\in \A(\Lambda_1)$ and $B\in \A(\Lambda_2)$ 
with $0 \in \Lambda_1$ and $0 \in \Lambda_2$
and $x$ satisfying $|x|-(d(\Lambda_1)+d(\Lambda_2))>0$,
\begin{eqnarray*}
\Vert \tau_x \alpha_t(A),B]\Vert
\leq&& 2 (N+1)^{d(\Lambda_1)+d(\Lambda_2)}
\Vert A\Vert \Vert B\Vert d(\Lambda_1)d(\Lambda_2)
\nonumber \\
&&
\mbox{exp}\{
-|t|(\frac{|x|-(d(\Lambda_1)+d(\Lambda_2))}{|t|} -2 V(\Phi))\}
\end{eqnarray*}%
holds.
\end{lemma}
This lemma guarantees the existence of a 
{\it finite group velocity} which is 
determined by Hamiltonian in the nonrelativistic setting.
Now we show the following lemma:
\begin{lemma}\label{th:start}
For an arbitrary function $f_T$ with the support $[-T,T]$ and 
satisfying $\int dt |f_T(t)|^2 <\infty$, the following relation 
holds:
\begin{eqnarray*}
\lim_{M \to \infty} \lim_{L \to \infty}
\int dt (\Omega, i[\hat{N}_{[-L,0]},\hat{H}_{[-M,M]}(t)] \Omega) f_T(t)
=\sqrt{2\pi} (\Omega,j_0 \Omega)\tilde{f}(0),
\end{eqnarray*}
where 
$\tilde{f_T}(\epsilon):=
\frac{1}{\sqrt{2\pi}}
\int dt f_T(t)e^{i\epsilon t}$ and $\hat{A}:=A-\omega(A)$
for $A\in{\cal A}$.
\end{lemma}
{\bf Proof:}
To esitamte the equation, let us consider the following 
quantity.
\begin{eqnarray}
&&(\Omega,[N_{[-L,0]},H_{[-M,M]}(t)]\Omega)
-(\Omega,[N_{[-L,0]},H_{[-M,M]}(0)]\Omega) \nonumber \\
&&
=
\int^t_0 ds (\Omega,[N_{[-L,0]},\alpha_s(\frac{d H_{[-M,M]}(u)}{du})]\Omega)
\nonumber \\
&&=-i
\int^t_0 ds (\Omega,[N_{[-L,0]},\alpha_s(
[H_{[-M,M]},H_{[-M-r+1,M+r-1]}])]\Omega)
\label{eq1}
\end{eqnarray}
The term $[H_{[-M,M]},H_{[-M-r+1,M+r-1]}])]$ expresses time derivative 
of energy contained in $[-M,M]$ and can be decomposed into in-going and 
out-going energy current.
\begin{eqnarray*}
i[H_{[-M,M]},H_{[-M-r+1,M+r-1]}]
=J_+-J_-,
\end{eqnarray*}
where 
$J_+ \in \A([M-r+2,M+r-1])$ is the in-going energy current at the 
left boundary   
and $J_- \in \A([-M-r+1,-M+r-2])$ represents the out-going
energy current at the right boundary.
Thus 
\begin{eqnarray}
(\ref{eq1})
=
\int^t_0 ds (\Omega,[N_{[-L,0]},
J_-(s)-J_+(s)]\Omega)
\label{eq2}
\end{eqnarray}
holds.
Now due to spacelike commutativity, $[N_{[-L,0]},J_+]=0$ holds 
and we obtain also for $J_-$, 
\begin{eqnarray*}
[N_{[-L,0]},J_-]
=&&-i [N_{[-L,0]},[H_{[-M,M]},H_{[-M-r+1,M+r-1]}]]
\nonumber \\
=&&i([H_{[-M,M]},[H_{[-M-r+1,M+r-1]},N_{[-L,0]}]]
\nonumber \\
&&
+[H_{[-M-r+1,M+r-1]},[N_{[-L,0]},H_{[-M,M]}]])
\nonumber \\
=&&
[H_{[-M,M]},-j_0]+[H_{[-M-r+1,M+r-1]},j]=0,
\end{eqnarray*}
where we used Jacobi identity for commutators.
To estimate (\ref{eq2}) we bound the deviation
\begin{eqnarray}
|(\Omega,[N_{[-L,0]},J_+(s)] \Omega)|
\leq \sum_{-L-M\leq z \leq -M}\Vert[n_z,J(s)]\Vert
\label{eq30}
\end{eqnarray}
where $\tau_{-L}(J_+)=:J$. Thanks to the Lemma \ref{th:gp},
it is bounded by
\begin{eqnarray}
(\ref{eq30})&&\leq
2(N+1)^{2r-1}\Vert n\Vert \Vert J\Vert(2r-1)
\sum_{-L-M\leq z\leq -M}
\mbox{exp}\{
-|s|(\frac{|z|-(2r-1)}{|s|}-2V(\Phi))\}
\nonumber \\
&&\leq
2(N+1)^{2r-1}\Vert n\Vert \Vert J\Vert(2r-1)
\frac{e^{-M}}{1-e^{-1}}e^{2r-1} e^{2|s| V(\Phi)}
\label{eq21}
\end{eqnarray}
Next we estimate the other term of (\ref{eq2}),
\begin{eqnarray}
&&|(\Omega,[N_{[-L,0]},J_-(s)]\Omega)|
=|(\Omega,[\alpha_{-s}(N_{[-L,0]},J_-]\Omega)|
\nonumber \\
&&=|(\Omega,[N_{[-L,0]},J_-]\Omega) +
(\Omega,\int^s_0 du [\alpha_{-u}(\frac{d\alpha_{-t}
(N_{[-L,0]})}{dt}|_{t=0},
J_-]\Omega)|
\nonumber \\
&&\leq
\int^s_0 du \Vert[\alpha_{-u}(\frac{d\alpha_{-t}(N_{[-L,0]})}{dt}|_{t=0}),
J_-]\Vert,
\label{eq3}
\end{eqnarray}
where we used $[N_{[-L,0]},J_-]=0$.
Now we can decompose as 
$\frac{d\alpha_{-t}(N_{[-L,0]})}{dt}|_{t=0}=j_{-L}-j_0$,
where $j_0\in \A([-r+2,r-1])$ and $j_{-L} \in \A([-L-r+1,-L+r-2])$.
These decompositions are used to obtain
\begin{eqnarray}
(\ref{eq3})
\leq 
\int^s_0 du \Vert[\alpha_{-u}(j).J_-]\Vert
+\int^s_0 du \Vert[\alpha_{-u}(j_-),J_-]\Vert.
\label{eq4}
\end{eqnarray}
By translating $J_-$ to the neighbourhood of origin, 
$J_0:=\tau_{M}(J_-) \in \A([-r+1,r-2])$, we can use 
lemma \ref{th:gp} to estimate the first term of (\ref{eq4}) as
\begin{eqnarray*}
&&\Vert [\alpha_{-u}(j),J_-]\Vert =\Vert[\tau_M\circ \alpha_{-u}(j),
J_0]\Vert
\nonumber \\
&&\leq
2 \Vert j\Vert \vert J_-\Vert
(N+1)^{4r-4}(2r-2)^2
\mbox{exp}\{-|u|(\frac{M-(4r-4)}{|u|}-2 V(\Phi))\}
\end{eqnarray*}
In the same manner we obatin the bound for socond term of (\ref{eq4}), 
\begin{eqnarray*}
&&\Vert[\alpha_{-u}(j_-),J_-]\Vert
=
\Vert[j_-, \alpha_u \circ \tau_{L-M} (J_0)]\Vert
\nonumber \\
&\leq&
 2 \Vert j\Vert \vert J_-\Vert
(N+1)^{4r-4}(2r-2)^2
\mbox{exp}\{-|u|(\frac{L-M-(4r-4)}{|u|}-2 V(\Phi))\}.
\end{eqnarray*}
Combination of the above estimates leads 
\begin{eqnarray}
(\ref{eq4}) \leq
\frac{e^{2V(\Phi)|s|}-1}{2V(\Phi)} 2\Vert j\Vert\Vert J_0\Vert
(N+1)^{4r-4} (2r-2)^2 
(e^{-M}+e^{-(L-M)})e^{4r-4}.
\label{eq22}
\end{eqnarray}
Thus, from (\ref{eq21}) and (\ref{eq22}), we obtain
\begin{eqnarray*}
&&|(\Omega,[N_{[-L,0]},H_{[-M,M]}(t)]\Omega)
-(\Omega,[N_{[-L,0]},H_{[-M,M]}]\Omega)|
\nonumber \\
&&\leq \int^t_0 ds
(\Vert [N_{[-L,0]},J_+(s)]\Vert +\Vert[N_{[-L,0]},J_-(s)]\Vert)
\nonumber \\
&&\leq Z_{M,L}(t),
\end{eqnarray*}
where 
\begin{eqnarray*}
Z_{M,L}(t)
:=&&2(N+1)^{2r-1}\Vert n\Vert \Vert J\Vert(2r-1)
\frac{e^{-M}}{1-e^{-1}}e^{2r-1}\frac{e^{2 V(\Phi)|t|}-1}{2 V(\Phi)}
\nonumber \\
&&
+ 2 \Vert j\Vert \Vert J_0\Vert
(N+1)^{4r-4}(2r-2)^2 
e^{4r-4}
\nonumber \\
&&
\{e^{-M}+e^{-(L-M)}
\frac{1}{2V(\Phi)}(\frac{e^{2V(\Phi)|t|}-1}{2 V(\Phi)}-|t| )\}.
\nonumber
\end{eqnarray*}
Finally integration with the function $f_T$ derives
\begin{eqnarray*}
&&|\int dt\{(\Omega,i [N_{[-L,0]},H_{[-M,M]}(t)]\Omega)
-(\Omega,i [N_{[-L,0]},H_{[-M,M]}]\Omega)\}f_T(t)|
\nonumber \\
&&\leq
\int dt|\{(\Omega,[N_{[-L,0]},H_{[-M,M]}(t)]\Omega)
-(\Omega,[N_{[-L,0]},H_{[-M,M]}]\Omega)\}||f_T(t)|
\nonumber \\
&&
=\int^{T}_{-T}dt |\{(\Omega,[N_{[-L,0]},H_{[-M,M]}(t)]\Omega)
-(\Omega,[N_{[-L,0]},H_{[-M,M]}]\Omega)\}||f_T(t)|
\nonumber \\
&&
\leq (\int dt |f_T(t)|^2 )^{1/2} 
(\int^{T}_{-T} dt |\{(\Omega,[N_{[-L,0]},H_{[-M,M]}(t)]\Omega)
-(\Omega,[N_{[-L,0]},H_{[-M,M]}]\Omega)\}|^2 )^{1/2}
\nonumber \\
&&
\leq (\int dt |f_T(t)|^2 )^{1/2} 
(2\int^{T}_{0} dt Z(t)^2 )^{1/2}
\nonumber \\
&&\leq
(\int dt |f_T(t)|^2 )^{1/2} 
\nonumber \\
&&
\{A(T)e^{-2M}+B(T)(e^{-2M}+e^{-2(L-M)}+2e^{-L})
+C(T)(e^{-2M}+e^{-L})\}^{1/2}
\end{eqnarray*}
where $A(T), B(T)$ and $C(T)$ do not depend upon $M$ and $L$.
Consequently we obtain the following:
\begin{eqnarray*}
\lim_{M \to \infty} \lim_{L \to \infty}
\int dt (\Omega, i[\hat{N}_{[-L,0]},\hat{H}_{[-M,M]}(t)] \Omega) f_T(t)
=\sqrt{2\pi} (\Omega,j_0 \Omega)\tilde{f}(0),
\end{eqnarray*}
The proof is completed.
\hfill {\bf Q.E.D.}
\\

This lemma gives a starting point for our
Goldstone-theorem-like argument which corresponds to 
(\ref{goldstone2}) in case of 
ordinary Goldstone theorem. 
Note that the ordering of limiting procedures,
$L \to \infty$ and $M\to \infty$, cannot be changed due to the 
definition of the current operator. In fact one can easily see that
if one takes $M\to \infty$ first, the left hand side of 
the above lemma vanishes.
\par
To study the property of energy momentum spectrum,
a proper correlation function should be investigated.
In physics literature, we often write the Hamitonian by the sumation of space
translated local hamiltonian at the origin. 
We define such a local object $h$ as follows.
\begin{definition}
Let $\Phi_1:=H_{\{0\}},\ \Phi_2:=H_{\{0,1\}}-(H_{\{0\}}+H_{\{1\}}),\ 
\Phi_3:=H_{\{-1,0,1\}}-(H_{\{0,1\}}+H_{\{-1,0\}})
\cdots,\Phi_{2m}:=H_{\{-m+1,\cdots,m\}}
-(H_{\{-m+1,\cdots,m-1\}}+
H_{\{-m+2,\cdots,m\}}),\
\Phi_{2m+1}:= H_{\{-m,\cdots,m\}}-
(H_{\{-m,\cdots,m-1\}}+H_{\{-m+1,\cdots,m\}})$ and define
\begin{eqnarray*}
h:=\sum_{s=1}^r \Phi_s \in {\cal A}([-r/2,r/2])
\end{eqnarray*}
with $h(y,t):=\tau_y \alpha_t(h)$. 
\end{definition}
Then the Hamiltonian can be written by the summation of 
space translated objects of $h$ as 
$H_{[-M,M]}=\sum_{y=-M+r}^{M-r}h(y,0)+C_{-M}+C_M$ where 
$C_{-M} \in {\cal A}([-M,-M+2r])$ and $C_M \in
{\cal A}([M-2r,M])$ represents the complmentary terms for
boundary.
\begin{definition}
To investigate the property of $E_{\omega}(dkd\epsilon)$ we 
define a "function" $\tilde{\rho}(k,\epsilon)$ as
\begin{eqnarray*}
\tilde{\rho}(k,\epsilon)dk d\epsilon
=(\Omega,i\hat{n}E_{\omega}(dk d\epsilon)\hat{h}\Omega).
\end{eqnarray*}
\end{definition}
Now it is the time to mention a theorem.
\begin{theorem}\label{th:main}
If $\omega$ preserves symmetry, i.e., $ \omega\circ \gamma_{\theta}
=\omega$,
\begin{eqnarray*}
2\pi i(\frac{\partial \tilde{\rho}(k,\epsilon)}
{\partial k}|_{k=0} -
\frac{\partial \tilde{\rho}(k,-\epsilon)}
{\partial k}|_{k=0})
=
(\Omega,j_0\Omega)\delta(\epsilon).
\end{eqnarray*}
holds.
\end{theorem}
{\bf Proof}\\
What we are interested in is the spectrum property with respec to $\omega$.
Its information is encoded in the left hand side of the above 
lemma \ref{th:start}.
To draw it we define functions $r_L$ and $s_M$ as 
\begin{eqnarray*}
r_L(x)&:=&1\ \mbox{for}\ -L\leq x \leq 0,\ \mbox{otherwise}\ 0
\\
s_M(x)&:=&1\ \mbox{for}\ -M\leq x \leq M,\ \mbox{otherwise}\ 0.
\end{eqnarray*}
By use of these objects and the spectrum decompotision of the 
space-time translation unitary 
operator $U(z,t):=\int e^{i(\epsilon t-kz)}E_{\omega}(dk d\epsilon)$, we obatin
\begin{eqnarray}
&&\int dt (\Omega, i[\hat{N}_{[-L,0]},\hat{H}_{[-M,M]}(t)] \Omega) f_T(t)
\nonumber \\
&&=\int dt \sum_{x}\sum_{y} 
r_L(x) s_M(y) (\Omega,i[\hat{n}(x,0),\hat{h}(y,t)]\Omega) f_T(t)
\nonumber \\
&&+
\int dt \sum_x r_L(x)
(\Omega, i[\hat{n}(x,0),C_{-M}(t)+C_M(t)] \Omega)f_T(t).
\label{eq5}
\end{eqnarray}
We denote Fourier transform of $\tilde{\rho}(k,\epsilon)$ as 
\begin{eqnarray*}
\rho(z,t):=\frac{1}{2\pi\sqrt{2\pi}}
\int d\epsilon \int^{\pi}_{-\pi} dk
\tilde{\rho}(k,\epsilon) e^{i(kz-\epsilon t)}
=\frac{1}{2\pi \sqrt{2\pi}}
(\Omega,i\hat{n} \hat{h}(-z,-t)\Omega),
\end{eqnarray*}
then we can write the equation (\ref{eq5}) as  
\begin{eqnarray}
(\ref{eq5})=&&
4\pi \sqrt{2\pi}\int dt \sum_z
Re(
 \rho(z,-t) (\sum_x r_L(x)s_M(x-z))) f_T(t)
\nonumber \\
&&+
\int dt \sum_x r_L(x)
(\Omega, i[\hat{n}(x,0),C_{-M}(t)+C_M(t)] \Omega)f_T(t).
\label{eq6}
\end{eqnarray}
Let us begin with an estimation of the second term of (\ref{eq6}),
\begin{eqnarray}
&&\int dt \sum_x r_L(x)
(\Omega, i[\hat{n}(x,0),C_{-M}(t)+C_M(t)] \Omega)f_T(t)
\nonumber \\
&&=\int dt \sum_x r_L(x)
(\Omega, i[\hat{n}(x,0),C_{-M}(t)] \Omega)f_T(t)
\nonumber \\
&&
+\int dt \sum_x r_L(x)
(\Omega, i[\hat{n}(x,0),C_M(t)] \Omega)f_T(t)
\label{eq11}
\end{eqnarray}
whose first term leads
\begin{eqnarray}
&&\int dt \sum_x r_L(x)
(\Omega, i[\hat{n}(x,0),C_{-M}(t)] \Omega)f_T(t)
\nonumber \\
&&
=\int dt \sum_x r_L(x)
(\Omega, i[\hat{n}(x,0),C_{-M}]\Omega) f_T(t)
\nonumber \\
&&
+\int dt \int^t_0 ds 
(\Omega, i[\alpha_s(\frac{d}{dt}\hat{N}_{[-L,0]}(-t)),
C_{-M}]\Omega) f_T(t)
\label{eq7}
\end{eqnarray}
Thanks to the symmetry of the state $\omega$ we can conclude 
the first term of the above equation (\ref{eq7}) vanishes and
can show that 
the second term also vanishes 
in the limit of $L \to \infty$ and $M\to \infty$ using 
the group-velocity lemma \ref{th:gp}.
The second term of (\ref{eq11})
can be shown to be also 
vanishing in the same limit in the same manner.
\par
Now the relation 
\begin{eqnarray*}
\lim_L \sum_x r_L(x)s_M(x-z)
=\left\{
\begin{array}{rl}
2M+1,& \quad z<-M \\
M+1-z,& \quad -M\leq z \leq M \\
0,& \quad M <z
\end{array}\right.
\end{eqnarray*}
is used to show the limiting value for $L$ to infinity as
\begin{eqnarray}
 \lim_{L \to \infty}
&&\int dt (\Omega, i[\hat{N}_{[-L,0]},\hat{H}_{[-M,M]}(t)] \Omega) f_T(t)
\nonumber \\
=&&4\pi\sqrt{2\pi}\int dt f_T(t)Re((\sum_{z<-M}\rho(z,-t)(2M+1))
\label{eq8} \\
+&&4\pi\sqrt{2\pi}\int dt f_T(t)Re((\sum_{-M\leq z\leq M}\rho(z,-t)(M+1))
\label{eq9} \\
+&&
4\pi\sqrt{2\pi}\int dt f_T(t)Re((-\sum_{-M\leq z \leq M}z \rho(z,-t))
\label{eq10}.
\end{eqnarray}
Next consider what will occur when $M$ is made infinity in the above
equation.
In the following, 
we show that (\ref{eq8}) and (\ref{eq9}) approach zero as $M \to \infty$.
Let us begin with (\ref{eq8}). 
\begin{eqnarray*}
(\ref{eq8})
&=&2 \int dt f_T(t)Re(
 \sum_{z>M} (\Omega,i \hat{n}\hat{h}(z,t)\Omega))(2M+1)
) \nonumber \\
&=&
i \int dt f_T(t)(\Omega,[\hat{n},\sum_{z>M}\hat{h}(z,t)]\Omega)(2M+1).
\end{eqnarray*}
Therfore due to Cauchy-Schwarz inequality,
one can obtain 
\begin{eqnarray*}
|(\ref{eq8})|
&\leq& \int dt |f_T(t)|
|(\Omega,[\hat{n},\sum_{z>M}\hat{h}(z,t)]\Omega)|(2M+1)
\nonumber \\
&\leq&
(2M+1)
(\int dt |f_T(t)|^2)^{1/2}
(\int^{T}_{-T} dt \Vert [\hat{n},\sum_{z>M}\hat{h}(z,t)]\Vert^2)^{1/2}.
\end{eqnarray*}
Since as for the integrand of the above equation,
the group-velocity lemma \ref{th:gp} is used to show  
\begin{eqnarray*}
\Vert [\hat{n},\hat{h}(z,t)]\Vert
\leq 2(N+1)^{r+1} \Vert \hat{n}\Vert \Vert \hat{h}\Vert
r exp[-|t|(\frac{|z|-(r+1)}{|t|}-2V(\Phi))]
\end{eqnarray*}
and 
\begin{eqnarray*}
\Vert [\hat{n},\sum_{z >M}\hat{h}(z,t)]\Vert
\leq  2(N+1)^{r+1} \Vert \hat{n}\Vert \Vert \hat{h}\Vert r
e^{r+1} e^{2|t|V(\Phi)}\frac{e^{-M}}{e-1}
\end{eqnarray*}
Thus finally we obtain
\begin{eqnarray*}
|(\ref{eq1})|
\leq
(\int dt|f_T(t)|^2)^{1/2}
(2M+1) 2(N+1)^{r+1} \Vert \hat{n}\Vert \Vert \hat{h}\Vert r
e^{r+1} \frac{e^{-M}}{e-1}
(\frac{e^{4V(\Phi)T} -1}{2V(\Phi)})^{1/2}
\end{eqnarray*}
which aproaches zero as $M\to \infty$.
\par
Next we estimate the equation(\ref{eq9}),
\begin{eqnarray*}
|(\ref{eq9})|
\leq 
(M+1) (\int dt |f_T(t)|^2)^{1/2}
\int^{T}_{-T} dt |(\Omega,[\hat{n},\sum_{-M}^M \hat{h}(z,t)]\Omega)|^2
)^{1/2}.
\end{eqnarray*}
The integrand of the above equation can be written by use of 
stationarity of $\omega$ as
\begin{eqnarray*}
(\Omega,[\hat{n},\sum_{-M}^M \hat{h}(z,t)]\Omega)
&=&(\Omega,[\hat{n},\sum_{-M}^M \hat{h}(z,0)]\Omega)
+\int^t_0 ds (\Omega,[\hat{n},\frac{d}{ds}\hat{H}_M(s)]\Omega)
\nonumber \\
&=&
\int^t_0 ds (\Omega,[\hat{n},\alpha_s(\frac{d\hat{H}_M}{dt}(0))]\Omega)
\nonumber \\
&=&
\int^t_0 ds (\Omega,[\hat{n},J_{-M}(s)-J_{M}(s)]\Omega).
\end{eqnarray*}
As before, decomposition into energy current terms 
\begin{eqnarray*}
\frac{d\hat{H}_M}{dt}=-[H_{[-M,M]},H_{[-M-r+1,M+r-1]}]
=J_{-M}-J_{M},
\end{eqnarray*}
where $J_{-M}\in\A ([-M-r+1,-M+r-2])$ and $J_M\in
\A ([M-r+2,M+r-1])$ leads
\begin{eqnarray*}
|(\Omega,[\hat{n},\sum_{-M}^M \hat{h}(z,t)]\Omega)|
\leq 
\int^t_0 ds \Vert [\hat{n},J_{-M}(s)]\Vert
+\int^t_0 ds \Vert [\hat{n}, J_M(s)]\Vert.
\end{eqnarray*}
And the following estimations which are 
obtained by direct use of group-velocity lemma \ref{th:gp} 
\begin{eqnarray*}
\Vert [\hat{n},J_{-M}(s)]\Vert
&\leq& 2(N+1)^{2r-1}\Vert n\Vert \Vert J_{-M}\Vert 2(r-1)
exp[(-|s|(\frac{M-(2r-1)}{|s|}-2V(\Phi))]
\nonumber \\
\Vert [\hat{n},J_{M}(s)]\Vert
&\leq& 2(N+1)^{2r-1}\Vert n\Vert \Vert J_{M}\Vert 2(r-1)
exp[(-|s|(\frac{M-(2r-1)}{|s|}-2V(\Phi))]
\nonumber
\end{eqnarray*}
with definition 
$max\{\Vert J_M\Vert,\Vert J_{-M} \Vert\}=:D >0$
leads,
\begin{eqnarray*}
|(\Omega,[\hat{n},\sum_{-M}^M \hat{h}(z,t)]\Omega)|
\leq 2(N+1)^{2r-1} \Vert n\Vert D 2(r-1)
e^{2r-1} e^{-M} \frac{e^{2V(\Phi)|t|}-1}{V(\Phi)}.
\end{eqnarray*}
Finally we obtain
\begin{eqnarray*}
&&(\int^{T}_{-T} dt|(\Omega,[\hat{n},\sum_{-M}^M
\hat{h}(z,t)\Omega)|^2)^{1/2}
\leq 2(N+1)^{2r-1} \Vert n\Vert D 
2(r-1)e^{2r-1} e^{-M}
\nonumber \\
&&
\frac{1}{V(\Phi)}
\sqrt{\frac{2}{V(\Phi)}}
(e^{2V(\Phi)T}-2e^{V(\Phi)T}+1+TV(\Phi))^{1/2}
\end{eqnarray*}
and can see 
\begin{eqnarray*}
\lim_{M \to \infty}(\ref{eq9}) =0
\end{eqnarray*}
holds.
Now we estimate the equation(\ref{eq10}).
Let
$\check{\rho}(k,t):=\sum_x \rho(x,t)e^{-ikx}$ be a Fourier
transform, then we obtain
\begin{eqnarray*}
-\sum_{-M}^M z\rho(z,-t)
&=&-\sum_{z=-M}^M \frac{1}{2\pi}\int^{\pi}_{-\pi}dk
\check{\rho}(k.-t)e^{ikz}z
\nonumber \\
&=&
-\sum_{-M}^M \frac{1}{2\pi}
\int dk \check{\rho}(k,-t) (-i)\frac{\partial}{\partial k}e^{ikz}
\nonumber \\
&=&
-\frac{i}{2\pi}\int dk(\frac{\partial }{\partial k} \check{\rho}
(k,-t))\sum_{-M}^M e^{ikz}.
\end{eqnarray*}
$\lim_M \sum_{-M}^M e^{ikz}=2\pi \delta(k)$ leads
\begin{eqnarray*}
\lim_{M \to \infty} (\ref{eq10})
&=&4\pi
\sqrt{2\pi} \int dt f_T(t)Re(-\frac{i}{2\pi} (\frac{\partial}{\partial k}
\check{\rho}(k,-t))|_{k=0})
\nonumber \\
&=&
4\pi \sqrt{2\pi}\int dt f_T(t) Re(-i\frac{\partial}{\partial k}
\check{\rho}(k,-t)|_{k=0})
\nonumber \\
&=&
i 2\pi \sqrt{2\pi}
\int d\epsilon
(\frac{\partial \tilde{\rho}(k,\epsilon)}
{\partial k}|_{k=0} -
\frac{\partial \tilde{\rho}(k,-\epsilon)}
{\partial k}|_{k=0})
\tilde{f}_T(\epsilon)
\end{eqnarray*}
Finally we obtain the following theorem:
\begin{eqnarray*}
2\pi i(\frac{\partial \tilde{\rho}(k,\epsilon)}
{\partial k}|_{k=0} -
\frac{\partial \tilde{\rho}(k,-\epsilon)}
{\partial k}|_{k=0})
=
(\Omega,j_0\Omega)\delta(\epsilon).
\end{eqnarray*}
The proof is thus completed.\hfill {\bf Q.E.D.}
\\

{\bf Proof of theorem \ref{th:th1}:}\\
If $\omega$ is symmetry breaking, original Goldstone theorem is applied 
to show $E_{\omega}(dkd\epsilon)$ to have singularity at the origin. 
If $\omega$ is not 
symmetry breaking, above theorem \ref{th:main} is applicable.
\hfill {\bf Q.E.D.}\\
\section{Conclusion and Outlook}
We considered states over one-dimensional infinite 
lattice which are stationary, translationally invariant and have non-vanishing
current expectations. The spectrum of space-time 
translation unitary operator with respect to such a state was investigated 
by use of Goldstsone-theorem-like technique and was shown to
have singularity at the origin $(k,\epsilon)=(0,0)$.
Although we employ the minimal definition of 
nonequilibrium steady state in the present paper,
physically more reasonable nonequilibrium steady state 
should be expected to give stronger result like poor decay 
of spatial correlation.
Alekseev et al. reported an interesting topic \cite{Fr}
which is related with this point.
\\

{\Large\bf Acknowledgement}
\\
I would like to thank Izumi Ojima and Yoshiko Ogata for fruitful discussions.
\appendix
\section{The proof of Lemma \ref{th:gp}}
{\bf Sketch of Proof:}
\\
According to theorem 6.2.11 in \cite{BR}, for $a,b \in 
\A[\{0\}]$,
\begin{eqnarray*}
\Vert [\tau_x \alpha_t (a),b]\Vert
\leq 2 \Vert a \Vert \Vert b\Vert
exp[-|t|(\frac{|x|}{|t|}-2V(\Phi)]
\end{eqnarray*}
holds. Our lemma is its finite range generalization.
Let $E_{\omega}(i_x,j_x),i_x,j_x=0,1,\cdots,N$ be a set of matrix
units for $\A[\{0\}]$. Each element $A$ and $B$ has a unique 
decomposition of the form,
\begin{eqnarray*}
A&=&\sum_{\{i_x\},\{j_x\}} C_A(\{i_x\},\{j_x\})\Pi_{x \in \Lambda_1}
E_{\omega}(i_x,j_x)
\nonumber \\
B&=&\sum_{\{i_x\},\{j_x\}} C_B(\{i_x\},\{j_x\})\Pi_{x \in \Lambda_2}
E_{\omega}(i_x,j_x)
\end{eqnarray*}
with coefficients $C_A, C_B \in {\bf C}$ satisfying 
$|C_A(\{i_x\},\{j_x\})|\leq \Vert A\Vert$
and $|C_B(\{i_x\},\{j_x\})|\leq \Vert B\Vert$.
Then the direct application of the theorem 6.2.11 in \cite{BR}
proves our theorem. 
\begin{eqnarray*}
\Vert \tau_x \alpha_t(A),B]\Vert
&\leq&
\sum_{\{i_x\},\{j_x\}} |C_A(\{i_x\},\{j_x\})|
\sum_{\{i'_x\},\{j'_x\}} |C_B(\{i'_x\},\{j'_x\})|
\nonumber \\
&&
\sum_{z \in \Lambda_1}\sum_{y\in \Lambda_2}
\Vert [\tau_x\alpha_t(E_{\omega}(i_z,j_z)),E_{\omega}(i'_y,j'_y)]\Vert
\nonumber \\
&=&
\sum_{\{i_x\},\{j_x\}} |C_A(\{i_x\},\{j_x\})|
\sum_{\{i'_x\},\{j'_x\}} |C_B(\{i'_x\},\{j'_x\})|
\nonumber \\
&&
\sum_{z \in \Lambda_1}\sum_{y\in \Lambda_2}
\Vert [\tau_{x+z-y}\alpha_t(E_{\omega}(i_0,j_0)),E_{\omega}(i'_0,j'_0)]\Vert
\nonumber \\
&\leq&
2(N+1)^{d(\Lambda_1)+d(\Lambda_2)}
\Vert A \Vert \Vert B\Vert
d(\Lambda_1)d(\Lambda_2)
\nonumber \\
&&
exp[-|t|(\frac{|x|-(d(\Lambda_1)+d(\Lambda_2))}{|t|}
-2V(\Phi))]
\nonumber
\end{eqnarray*}
\hfill {\bf Q.E.D.}

\end{document}